\documentclass[%
reprint,
 nofootinbib,
aps,
prl,
]{revtex4-2}

\usepackage{graphicx}
\usepackage{physics}
\usepackage{dcolumn}
\usepackage{kpfonts}
\usepackage{bm}
\usepackage{hyperref}
\begin{document}

\preprint{APS/123-QED}

\title{Optomechanical disk resonator in the quantum ground state of motion}

\author{Andrea Barbero$^{1}$} 
\altaffiliation{These authors contributed equally to this work.}
\author{Samuel Pautrel$^{1}$} 
\altaffiliation{These authors contributed equally to this work.}
\author{Bertrand Evrard$^1$}%
\author{Jérémy Bon$^1$}%
\author{Romain Dezert$^1$}%
\author{Aristide Lemaître$^2$}%
\author{Adrien Borne$^1$}%
\author{Ivan Favero$^{1,}$}%
\email{ivan.favero@u-paris.fr}
\affiliation{%
 $^1$Université Paris Cité, CNRS, Matériaux et Phénomènes Quantiques, 75013 Paris, France\\
 $^2$Université Paris-Saclay, CNRS, Centre de Nanosciences et de Nanotechnologies, 91120 Palaiseau, France 
}%

\date{\today}

\begin{abstract}
Although they have enabled several advances in the field of optomechanics, optomechanical disk resonators have not yet been operated in the quantum regime. We present the first experimental demonstration of an optomechanical disk resonator prepared in the quantum ground state. With a gigahertz frequency, the mechanical breathing mode of the investigated semiconductor disk reaches a level of excitation below a single phonon when cooled in a dilution refrigerator. We quantify the phonon occupancy of the mechanical mode by performing Brillouin sideband spectroscopy: a conical optical fiber is evanescently coupled to the disk optical whispering-gallery mode, and Stokes and anti-Stokes photons scattered by phonon emission and absorption are counted on a single-photon detector. We measure a suppression of the absorption process corresponding to a phonon occupancy of $0.66\pm0.20$. We experimentally investigate the mechanisms ruling laser-induced heating, which limits the lowest measurable phonon occupancy, and notably witness an extra-cavity heating effect.
\end{abstract}

\maketitle
The coherent control of mechanical resonators using light, enabled by optomechanical interactions \cite{favero_optomechanics_2009,aspelmeyer_cavity_2014}, plays a growing role in the development of quantum technologies \cite{barzanjeh_optomechanics_2022}. Mechanical resonators are indeed a very promising platform for several applications such as quantum memories \cite{julsgaard_experimental_2004} (mechanical vibration can store information \cite{wallucks_quantum_2020,chu_creation_2018} and can reach a second lifetime for a GHz vibration \cite{maccabe_nano-acoustic_2020}), quantum transducers \cite{kimble_quantum_2008} (mechanical conversion of quantum information between microwave and telecom photons \cite{vainsencher_bi-directional_2016,arnold_converting_2020,mirhosseini_superconducting_2020,jiang_optically_2023,blesin_bidirectional_2024,van_thiel_optical_2025,brubaker_optomechanical_2022,zhao_quantum-enabled_2025}), and quantum sensing \cite{degen_quantum_2017} (through the exploitation of squeezed \cite{pirkkalainen_squeezing_2015,marti_quantum_2024} or non-classical \cite{pan_realization_2025} states to enhance sensitivity).\\
A common prerequisite for exploiting mechanical resonators in the quantum regime is their preparation in the ground state of motion \cite{poot_mechanical_2012}. This challenging regime has been reached using different cooling techniques, such as sideband and cavity-assisted \cite{teufel_sideband_2011,noguchi_ground_2016,peterson_laser_2016,qiu_laser_2020,noguchi_ground_2016,delic_cooling_2020,galinskiy_phonon_2020,deplano_high_2025}, feedback \cite{tebbenjohanns_quantum_2021,seis_ground_2022} and cryogenic cooling \cite{oconnell_quantum_2010,meenehan_pulsed_2015,cattiaux_macroscopic_2021,ramp_elimination_2019,doeleman_brillouin_2023,chu_quantum_2017,arrangoiz-arriola_resolving_2019,mayor_high_2025}. Among the diverse geometries of resonators employed in these early demonstrations, a handful of devices only ventured into concrete sensing experiments, albeit in the classical regime \cite{bagci_optical_2014,ahn_ultrasensitive_2020,halg_membrane-based_2021,chowdhury_membrane-based_2023,liang_yoctonewton_2023,han_feedback-control_2023,salimi_squeeze_2024}. This status reflects the difficulty of optimizing a resonator both to operate in the quantum regime and to be a real sensor. On their side, optomechanical disk resonators \cite{ding_high_2010,sun_high-q_2012} already have a consolidated series of sensing applications to their credit. They indeed showed their capacity to sense the action of optical \cite{guha_force_2020} and atomic forces \cite{allain_optomechanical_2020}, magnetic fields \cite{forstner_cavity_2012,yu_optomechanical_2016,hu_picotesla-sensitivity_2024}, the mass of individual nanoparticles \cite{sbarra_multimode_2022,liu_sub-pg_2013}, the viscoelastic interactions at play with a liquid \cite{neshasteh_optomechanical_2025} and even the vibration of a single bacterium \cite{gil-santos_optomechanical_2020}. Their simple and highly symmetric geometry presents many assets for sensing : it facilitates signal interpretation, allows very high optical quality factor \cite{borselli_beyond_2005,guha_surface-enhanced_2017}, and lends itself to dense on-chip integration \cite{lamberti_real-time_2022}. Despite these merits for sensing, the operation of an optomechanical disk resonator in the quantum regime remains to be demonstrated.

Here we report on the preparation of a gigahertz-frequency optomechanical disk resonator close to its ground state of motion, with mean phonon occupancy $0.66 \pm 0.20$. The disk is measured in a dilution cryostat with base temperature of 11 mK and the occupancy of the mechanical mode is evaluated by counting the single photons scattered into the Brillouin motional sidebands through optomechanical interaction \cite{meenehan_pulsed_2015,cohen_phonon_2015}. At 4 K, the count rates are proportional to the injected optical power P, following canonical expectations. At millikelvin temperatures, the evolution is at variance superlinear, revealing heating by the injected light. The power scaling of this heating, investigated under varying measurement sequences, is compatible with the concomitant existence of a fast linear intracavity absorption and a slow extracavity heating process. This observation clarifies the optimal conditions of operation of optomechanical resonators for future sensing experiments in the quantum regime.

\begin{figure}[h!]
    \centering
    \includegraphics[scale=0.8,page=1,trim={1cm 14cm 10.5cm 1cm},clip]{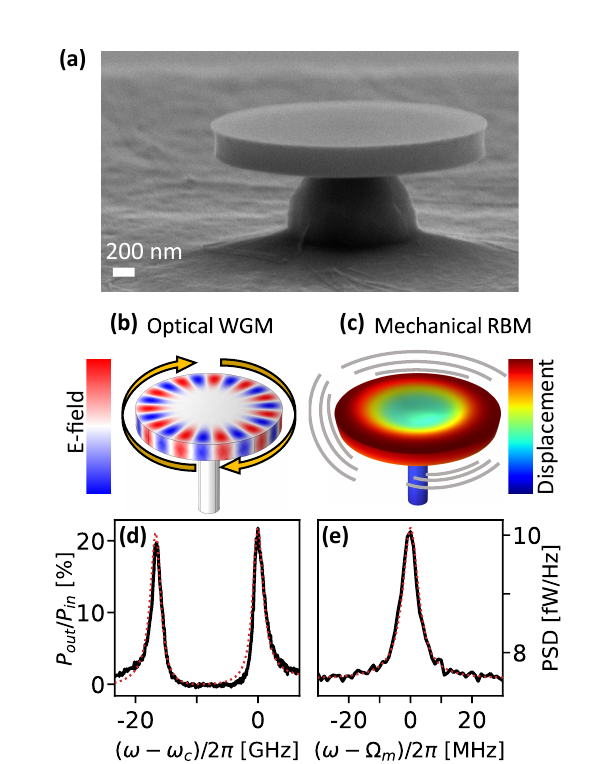}
    \caption{(a) Scanning electron microscope image of an optomechanical GaAs disk resonator. (b) Radial component of the electric field of the optical WGM with azimuthal number $m=22$ and radial number $p=1$. (c) Norm of the displacement field of the fundamental mechanical RBM. (d) Doublet optical spectrum \cite{borselli_beyond_2005} measured in reflection through the input fiber and (e) mechanical spectrum measured at room temperature around the RBM resonance frequency $\Omega_m/2\pi = 1.085$ GHz. Solid black lines and dashed red lines in (d) and (e) correspond to measured signals and Lorentzian fits, respectively.}
    \label{Fig1}
\end{figure}

Our resonator consists in a 1.3 $\mu m$-radius, 320 nm-thick gallium arsenide (GaAs) disk mounted on a $\approx500$ nm-radius, 1.8 $\mu$m-tall aluminum gallium arsenide (AlGaAs) pedestal (FIG. \ref{Fig1} (a)). Its surface is passivated in order to limit optical absorption \cite{guha_surface-enhanced_2017,najer_suppression_2021,supplementary}. In numerical simulations, the disk hosts two degenerate optical whispering-gallery modes (WGM) at resonance frequency $\omega_c/2\pi = 196.78$ THz ($\lambda_c=1523.49$ nm), illustrated in FIG. \ref{Fig1} (b), and a mechanical radial breathing mode (RBM FIG. \ref{Fig1} (c)) at resonance frequency $\Omega_m/2\pi = 1.085$ GHz. In practice, the two degenerate clockwise and counter-clockwise WGMs are coupled by backscattering and give rise to two frequency-split eigenmodes associated to standing waves, visible in the optical spectrum at 4 K (FIG. \ref{Fig1} (d)). The spectrum is obtained by evanescent optical coupling to the disk, performed by approaching a single-sided conical fiber to it \cite{pautrel_efficient_2024}. The mechanical spectrum of the Brownian motion at room temperature is obtained from the radio-frequency fluctuations of the output light, as shown in FIG. \ref{Fig1} (e). The optical intrinsic loss rate and external coupling rate of the WGM, and the mechanical damping rate of the RBM, are extracted from these data: $(\kappa_{i},\kappa_{e},\gamma_{m})/2\pi = (1585, 480, 6)$ MHz, respectively, resulting in loaded optical and mechanical quality factors $Q_o\approx10^5$ and $Q_m \approx 180 $. The WGM and RBM are coupled at the single photon optomechanical coupling rate $g_0/2\pi = 220 $ kHz, calculated using finite-element method (FEM) simulations. The modest value of $Q_m$ is associated to a relatively wide pedestal, chosen to facilitate thermal coupling to the substrate. The sample is located on a sample holder thermalized to the mixing chamber plate of a dry dilution cryostat. Cryogenic piezo positionners allow the motion of the optical conical fiber, monitored with a imaging system involving a cryogenic microscope objective, a series of windows on each radiation shield and a camera outside the cryostat.

The sideband thermometry measurement is performed by driving the optomechanical system with a sequence of optical pulses, spectrally red- or blue-detuned by one mechanical frequency with respect to the optical mode resonance. These pulses are generated by two independent continuous-wave lasers whose light passes through acousto-optic modulators. The pulse sequence consists of a red pulse and a blue pulse, each lasting $\tau_{p}=$ 4 $\mu s$ and separated by a delay $\tau$ of 1 $\mu s$, repeated for a duration of 2.5 seconds. The light exiting from the cavity is spectrally filtered around the optical resonance to detect only the optomechanically-scattered photons, which are counted on a SPD, as illustrated in FIG. 2 (a). The filtering system consists in two tunable Fabry-Perot cavities in series spectrally aligned, both with a bandwidth of 10 MHz and a free spectral range of 5 GHz \cite{supplementary}. This cascade of filters allows for a transmission efficiency of $\approx50\%$ and an extinction ratio of $\approx80$ dB at 1.1 GHz from resonance \cite{galinskiy_phonon_2020}. It remains stable for about 2.5 seconds, matched to the pulse sequence duration. Beyond this duration it spectrally drifts and its transmission drops by $15\%$. The SPD is a superconducting nanowire cooled down to 2K and with an applied external current bias that allows for the best dark count-limited signal-to-noise ratio (SNR), resulting in $C_{dark}=11$ Hz dark count rate and detection efficiency $\approx85\%$. Since the employed disk optical mode is a standing superposition of clockwise and counter-clockwise waves, it radiates part of its energy back into our single-sided conical fiber. Looking at these photons travelling backwards allows for preliminar rejection of the laser light. Indeed only the laser photons that entered the disk mode have to be filtered out, which improves the SNR.

\begin{figure}
    \centering
    \includegraphics[scale=0.9,page=2,trim={1.5cm 13.5cm 10.5cm 1cm},clip]{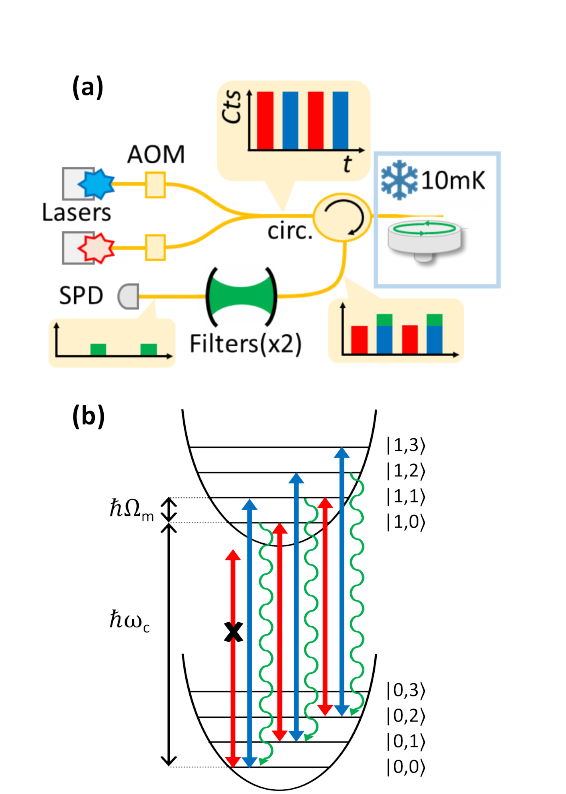}
    \caption{(a) Sketch of the experimental setup. A sequence of laser pulses, alternately red- and blue-detuned with respect to the optical mode resonance, probe the disk resonator placed in the dilution refrigerator. Brillouin-scattered single photons at the optical resonance frequency, resulting from the optomechanical interaction, are detected with a SPD, while the probe photons are spectrally rejected. (b) Eigenenergies of the system $\ket{n_a,n_b}$, not to scale ($n_a$ and $n_b$ the photon and phonon numbers, respectively). Photons spectrally detuned by one quantum of vibration from the optical resonance can drive Stokes/anti-Stokes processes (blue/red arrows), resulting in single photons scattered at the optical resonance (green arrows) at probabilities proportional to $n_b+1$ (Stokes) or $n_b$ (anti-Stokes).}
    \label{Fig2}
\end{figure}

The rate of detection of the optomechanically-scattered photons under blue-(red-) detuned drive $\Gamma_{b\left(r\right)}$ is directly related to the mean phonon occupancy $\langle n_b\rangle$ \cite{meenehan_pulsed_2015}:
\begin{align}
    \Gamma_b&=\eta_{tot}\frac{4\kappa_{e}}{\kappa_{tot}^2}g_0^2\langle n_a\rangle\left(\langle n_b\rangle+1\right)\label{eqn:gamma_b}\\
    \Gamma_r&=\eta_{tot}\frac{4\kappa_{e}}{\kappa_{tot}^2}g_0^2\langle n_a\rangle\langle n_b\rangle\label{eqn:gamma_r}
\end{align}
where $\langle n_a\rangle$ is the mean intracavity photon number, $\kappa_{tot}=\kappa_e+\kappa_i$, and $\eta_{tot}$ the total detection efficiency. The count rate on the SPD is $\Gamma_{SPD}=\Gamma_{\left(b,r\right)}+\Gamma_{dark}+\Gamma_{pump}$, where $\Gamma_{dark}$ is the dark count rate and $\Gamma_{pump}$ is the probe leakage count rate, the latter being negligible thanks to our filtering system combined with the backward detection configuration.
From the measured count rates we can thus infer the mean phonon occupancy of the mechanical mode as:
\begin{align}
    \langle n_b\rangle&=\frac{1}{\Gamma_b/\Gamma_r-1}.
\end{align}
When the system sits in the ground state of motion, the anti-Stokes optomechanical process is forbidden and $\Gamma_r=0$, as shown in FIG. \ref{Fig2} (b).

\begin{figure}
    \centering
    \includegraphics[scale=0.8,page=2,trim={1.5cm 2.5cm 10.5cm 14.5cm},clip]{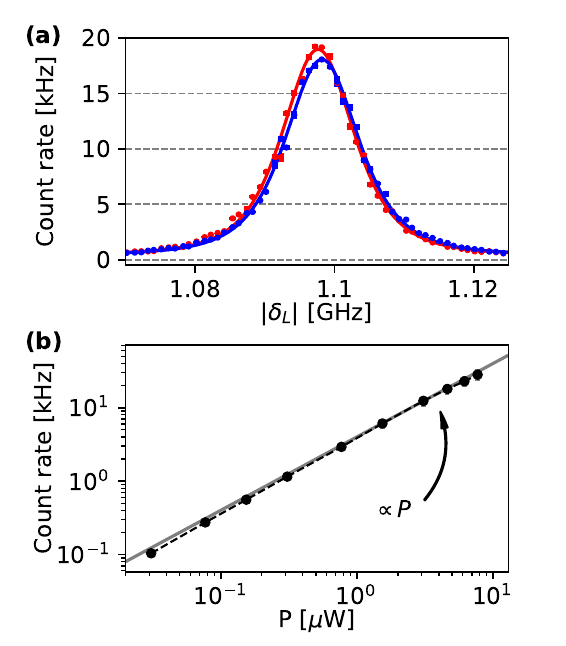}
    \caption{Measurements at 4K temperature. (a) Brillouin-scattered single-photon count rates as a function of the optical probe detuning to the optical resonance frequency, under pulse-on input optical power $P$=7.6 $\mu$W. (b) Amplitudes of the mean value of Stokes and anti-Stokes sideband count rates when $\delta_L=\pm\Omega_m/2\pi$, as function of optical power $P$, for two different datasets, showing a close-to-linear behaviour. The inferred single-photon optomechanical coupling rate is $g_0/2\pi=212\pm14$ kHz. For both panels, we subtracted the dark count rate.}
    \label{Fig3}
\end{figure}

We benchmark our setup at 4K, where the thermal phonon occupancy of the disk RBM is large enough ($\langle n_b\rangle\approx80$) to make the asymmetry between $\Gamma_b$ and $\Gamma_r$ practically undetectable. We therefore probed the disk at constant input optical mean power (6.08 $\mu$W) for different detunings of the probe laser with respect to the optical resonance, and our measurements shown in FIG. \ref{Fig3} (a) reveal a quasi-perfect equality between the anti-Stokes and Stokes optomechanical processes (red and blue, respectively). The measured curves, which are the convolution of the filters response and the mechanical mode, are peaked at the detuning $\delta_L=\pm\Omega_m/2\pi$. We investigated the evolution of the count rates with optical input power, shown in FIG. \ref{Fig3} (b). The evolution is linear, consistent with equations (\ref{eqn:gamma_b},\ref{eqn:gamma_r}), which shows that laser-induced heating has negligible consequences at this cryostat temperature. After a proper calibration of the intracavity photon number, we extracted the optomechanical vacuum coupling rate using equations (\ref{eqn:gamma_b},\ref{eqn:gamma_r}), finding $g_0/2\pi=212\pm14$ kHz, in good agreement with the simulated value.

\begin{figure}[!ht]
    \centering
    \includegraphics[scale=0.8,page=3,trim={1cm 6cm 10.5cm 1cm},clip]{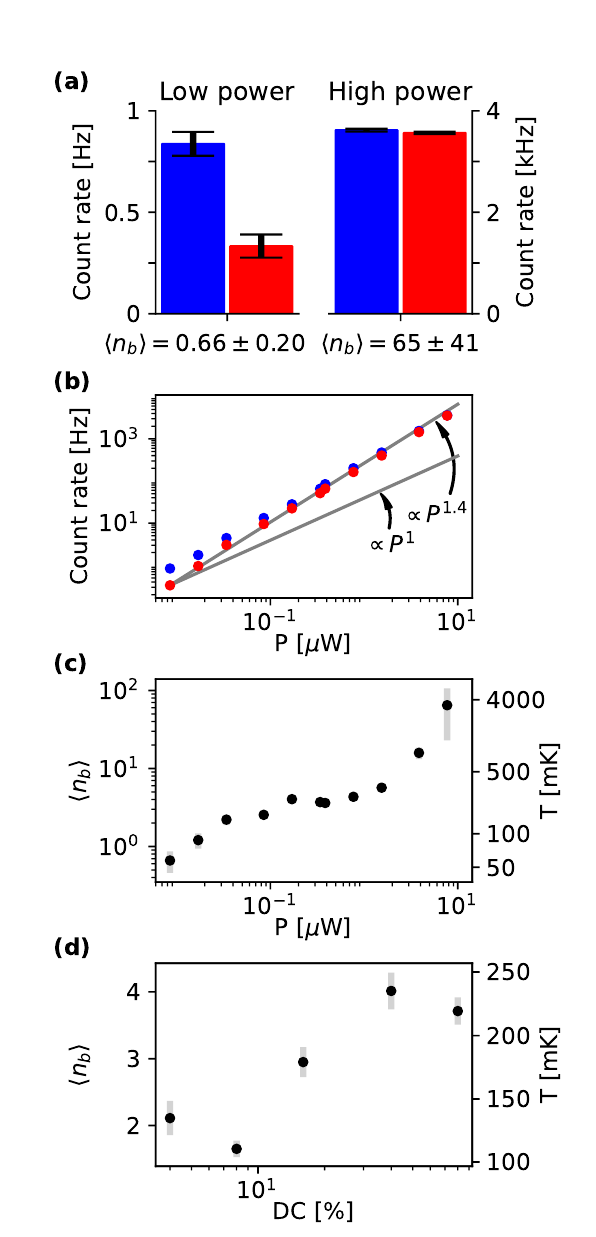}
    \caption{Measurements at millikelvin temperatures. (a) Brillouin-scattered single-photon count rates under blue- and red-detuned driving at $\delta_L=\pm\Omega_m/2\pi$ at low (8.5 nW) and high (7.7 $\mu$W) optical mean power. (b) Stokes and anti-Stokes sideband amplitudes. (c) Inferred mechanical occupancy and corresponding effective modal temperature as a function of the probe pulse-on optical power $P$. (d) Inferred mechanical occupancy and modal temperature as a function of the duty cycle $DC$ with constant $\langle n_a\rangle$. For (a) and (b) panels, we subtracted the dark count rate.}
    \label{Fig4}
\end{figure}

We then cooled down the system at $\approx11$ mK, read-out on a thermometer placed on the mixing chamber plate. At this temperature the thermal phonon occupancy of the disk RBM is expected to be $\approx0.009$ according to the Bose-Einstein statistics. We measured the Stokes and anti-Stokes count rates when optically pumping at $\delta_L=\pm\Omega_m/2\pi$, for two different laser probe mean powers, as shown in FIG. \ref{Fig4} (a). When the mean power is 7.7 $\mu$W, the laser-induced heating results in a large phonon occupancy and the two Brillouin processes have nearly equal probability. In contrast, with a mean power of 8.5 nW, we measured a clear asymmetry between the two processes, which leads us to evaluate a mean phonon occupancy $\langle n_b\rangle=0.66\pm0.20$. This corresponds to a probability to be in the vacuum state close to $60\%$.

We repeated this sideband thermometry measurement as a function of the pulse-on laser power $P$, for the same pulse sequence as above. As illustrated in FIG. \ref{Fig4} (b), we observed a clear superlinear evolution of the count rates, with a power scaling close to $P^{1.4}$. This is a signature of laser-induced heating of the disk, further confirmed by looking at the inferred mean phonon occupancy and associated modal temperature in FIG. \ref{Fig4} (c), which both increase with $P$. Thermalization effects were reported to be of crucial importance for the phononic occupation of some previous experimental demonstrations \cite{meenehan_pulsed_2015,mayor_high_2025}. To investigate the timescales involved in our heating process, we performed a pump-probe measurement where an intense pump pulse heats the device, and a weak probe pulse measures the resulting count rate. We varied the delay between pump and probe from 100 ns to 20 $\mu s$, while keeping the average power constant. FIG. 7 of \cite{supplementary} reports that the resulting count rate is constant over all explored delays, showing that heating produced by the pump pulse relaxes faster than 100 ns, compatible with fast thermal relaxation of the disk, or/and slower than 20 $\mu s$, compatible with slow processes external to the disk. We then varied the duty cycle $DC=\tau_{p}/(\tau+\tau_{p})$ of the thermometry pulse sequence, by varying $\tau$ while keeping $P$ constant (therefore the intracavity photon number $\langle n_a\rangle$ constant during the pulses, given the $\sim$ns transient dynamics of the optical cavity). When $DC$ increases, FIG. \ref{Fig4} (d) shows that the phonon occupancy evaluated during the pulses increases as well. Given the $<100$ ns relaxation of the intracavity heating, this observed behaviour is a signature of the existence of an extracavity heating process, with a timescale long compared to $\tau$ and $\tau_{p}$. The ensemble of our data is hence compliant with the concomitant existence of fast intracavity heating ($<$100 ns) and slow extracavity heating ($>$20 $\mu s$).

To conclude, we experimentally demonstrated the preparation of an optomechanical disk resonator close to its ground state of motion, through cryogenic cooling and optical sideband thermometry. Although we measured a mean phonon occupancy of $0.66\pm0.20$, we also showed that our system is still limited by laser-induced heating. Reducing the optical power even further is achievable in our setup if collecting both transmission and reflection signals from the coupling fiber, and if placing the SPD at millikelvin in order to reduce the dark counts. With the present report, we place optomechanical disk resonators in the quantum regime, which, in combination with the advantages this platform presents for sensing, opens to a new set of experiments to be performed.

\bibliography{biblio}

\clearpage
\begin{widetext}
\section*{Supplementary Material}

%
%




\section{Fabrication of Gallium Arsenide disks}
The nanofabrication of optomechanical disks is performed using a bulk GaAs wafer on which is epitaxially grown a 1.8 $\mu$m-thick layer of Al\textsubscript{0.8}Ga\textsubscript{0.2}As (to form the pedestal) and then a 320 nm-thick top layer of GaAs (to form the disk). The subsequent top-down fabrication process is illustrated in Figure \ref{fig:Fab} and can be summarized into the three following main steps: 
\begin{itemize}
    \item The definition of GaAs/AlGaAs cylinders with electron beam lithography and dry vertical etching based on Ar/SiCl$_4$ inductively coupled plasma (ICP);
    \item The formation of the "mesa", a structure raised above the substrate on which the disks lie elevated. This is done by protecting the cylinders using a photolithography process and performing a wet etching in an acid solution composed of H$_3$PO$_4$ : H$_2$O : H$_2$O$_2$ (1:1:1);
    \item The pedestal is formed by performing a wet etching in HF (1.25$\%$) at 4°C, which selectively etches the AlGaAs layer without etching the GaAs.
\end{itemize}

\begin{figure}[!h]
    \centering
    \includegraphics[width=0.8\textwidth]{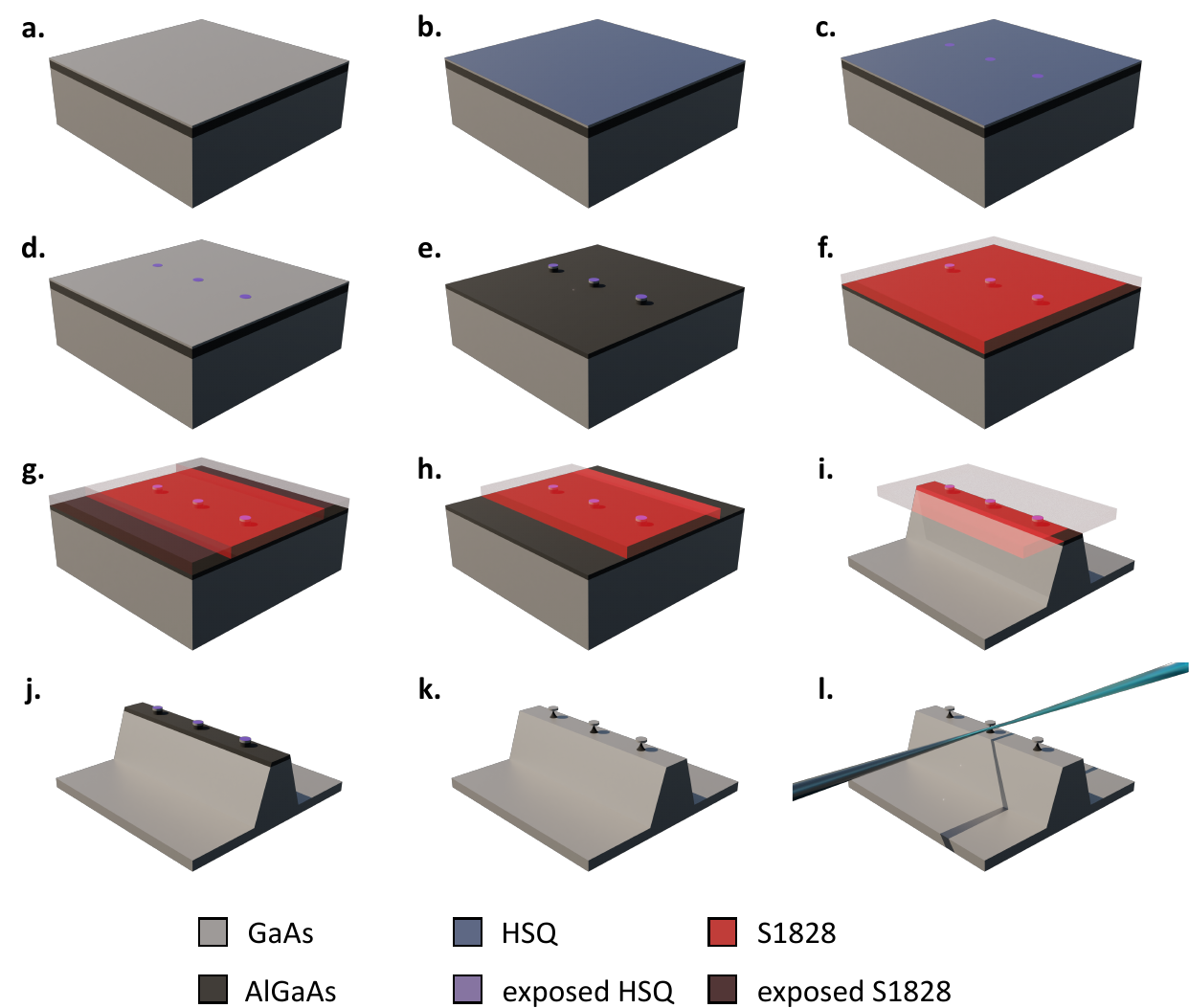}
    \caption{(Not to scale) Representation of the GaAs disks fabrication process. \textbf{a.} Cleaned wafer; \textbf{b.} spin coating of HSQ resist; \textbf{c.} electron beam lithography in the Scanning Electron Microscope (SEM); \textbf{d.} development of the resist to remove the unexposed HSQ; \textbf{e.} vertical dry etching with the ICP; \textbf{f.} spin coating of S1828 resist; \textbf{g.} UV photolithography; \textbf{h.} development of the resist to remove the exposed S1828; \textbf{i.} mesa wet etching with $\text{H}_3\text{PO}_4$; \textbf{i.} hot bath of SVC14 to remove unexposed S1828; \textbf{k.} HF underetching; \textbf{l.} fabricated devices with conical fibers for coupling light to the disks.}
    \label{fig:Fab}
\end{figure}

In Figure \ref{fig:Disks_on_mesa} we report a scanning electron microscope (SEM) micrograph representing a series of disk resonators lying on the mesa.

\begin{figure}[!h]
    \centering
    \includegraphics[width=0.8\textwidth]{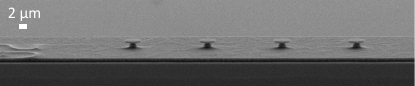}
    \caption{SEM micrograph showing a series of disks and the mesa from a side view.}
    \label{fig:Disks_on_mesa}
\end{figure}

The last step of the fabrication consists of a surface treatment with the purpose of enhancing the intrinsic optical quality factors of the devices by reducing the absorption losses in the amorphous surface reconstruction layer of GaAs. This layer is firstly etched with diluted NH$_4$OH (6$\%$), which only attacks oxidized layers, and the sample is then dipped in diluted (NH$_4$)$_2$S ($20\%$), which "sulfurizes" the surface to prevent the very fast oxidation of GaAs ($\approx$ms). The sulfur layer on the surface is very fragile, but gives us enough time to perform an Atomic Layer Deposition (ALD) of 10 nm of Al$_2$O$_3$, which finalizes the surface of our resonators. Figure \ref{fig:ALD} shows the intrinsic optical quality factors $Q_i=\omega_c/\kappa_i$ of different measured disks before and after the ALD surface passivation. The ALD treatment corresponds to a clear enhancement of the intrinsic optical quality factor, if it is not sidewall-roughness limited.  

\begin{figure}[!h]
    \centering    \includegraphics[width=0.4\textwidth]{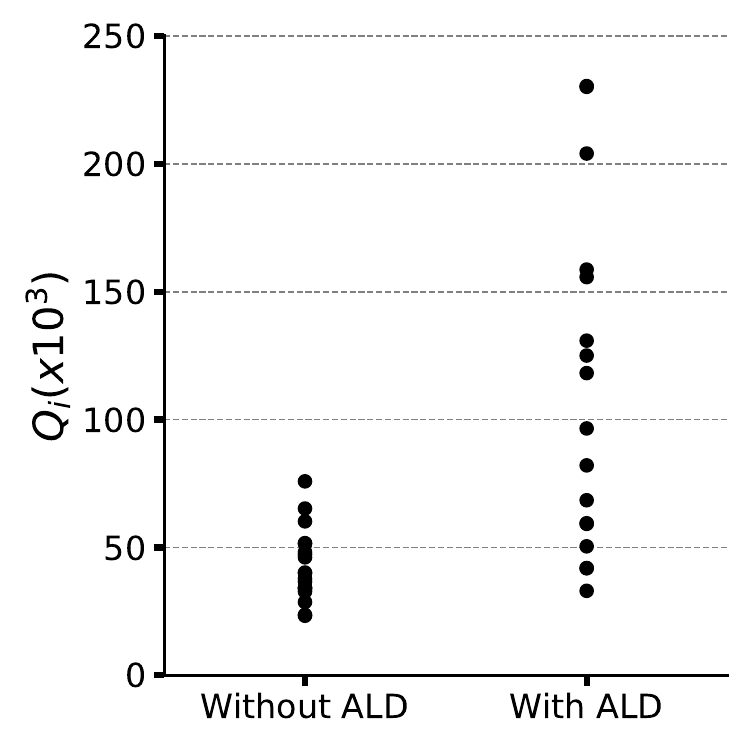}
    \caption{Measured intrinsic optical quality factors of optomechanical GaAs disks with (right) and without (left) the ALD-based surface treatement.}
    \label{fig:ALD}
\end{figure}

\newpage
\section{Experimental setup and methods}

\begin{figure}[!h]
    \centering   
     \includegraphics[width=0.85\textwidth]{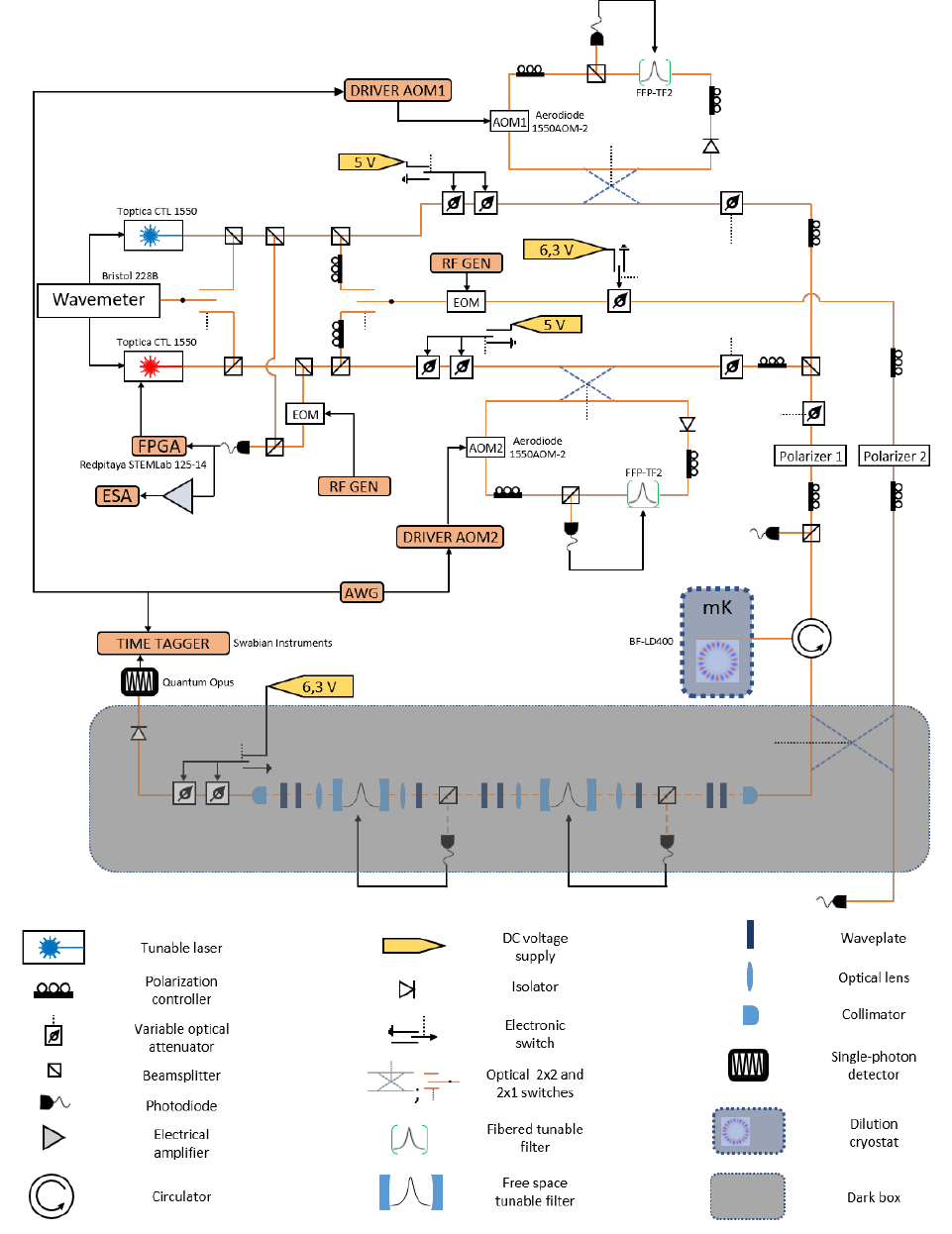}
    \caption{Photon-counting setup implemented for sideband thermometry measurements. The device is here probed in reflection. AOM: acousto-optic modulator; EOM: electro-optic modulator; ESA: electronic spectrum analyser; FPGA: field programmable gate array; AWG: arbitrary waveform generator.}
    \label{fig:PC_setup}
\end{figure}

\newpage
The experimental setup can be divided into three main sections:
\begin{itemize}
\item The preparation of the optical temporal sequence, which involves two continuous-wave tunable lasers (Toptica CTL 1550) that are each properly locked to a target wavelength and to a target power, and whose light is shaped into optical pulses through acousto-optic modulators (Aerodiode 1550AOM-2); 
\item A cryogenic environment hosting the optomechanical device, which involves a dilution cryostat (Bluefors BF-LD400), cryogenic piezo positionners (Attocube), and an in-situ imaging system to enable optical fiber coupling to the device;
\item A detection stage, which involves a cascade of two home-made optical filters and a single-photon detector (SPD) (Quantum Opus), put in the dark to avoid environmental light contamination of the signal.
\end{itemize}
All the optical elements are fibered except the detection stage involving the optical filters.\\

The wavelength stabilization is performed using a master-slave technique. With a proportional–integral–
derivative (PID) control, we lock the absolute value of the wavelength of one laser (\textit{master}), read on a wavemeter (Bristol 228B). A second laser (\textit{slave}) is modulated at the frequency $2\Omega_{target}-\Delta$. The two lasers are then mixed together on a fast photodiode, which results in different interference beatnotes: $2\Omega_{target}$ between the two pump lasers, $2\Omega_{target}-\Delta$ between the slave pump laser and its sidebands, $\Delta$ between the master pump laser and one slave pump laser sideband, and $3\Omega_{target}-\Delta$ between the master pump laser and the other slave pump laser sideband. This signal is low-pass filtered thanks to the bandwidth of our FPGA (Redpitaya STEMLab 125-14) to keep only the component oscillating at $\Delta$. It is then mixed with a local oscillator at the same frequency $\Delta$ to generate a DC feedback signal, which is part of a second PID that maintains the spectral position of the slave laser to maximize this DC signal. This maintains the detuning between the slave-laser sideband and the master laser constant at $\Delta$, and thus the detuning between master and slave at $2\Omega_{target}$. Moreover, both lasers pass through a tunable fibered band-pass filter (FFP-TF2 Luna Inc.) spectrally centered around their emission wavelengths in order to reject their high frequency phase noise. \\

The filtering stage, consisting of a series of two homemade tunable Fabry-Perot cavities, is properly designed to ensure an extinction ratio of 80 dB at 1.1 GHz from the optical resonance when the two cavities are spectrally aligned. One of the mirrors composing the cavity is mounted on a piezoelectric actuator, which allows modifying the cavity length and thus the resonance frequency. Both cavities have been designed to have a bandwidth $\gamma_f=10$ MHz; their free spectral ranges (FSR) are 5 GHz and 5.8 GHz. Figure \ref{fig:Filters_carac} shows the spectral characterization of one cavity, performed by recording on a photodiode the reflected light when a pump laser with a fixed wavelength is sent to it and when varying the applied voltage on the piezoelectric actuator (after a proper calibration of the frequency/voltage).

\begin{figure}[!h]
    \centering
    \includegraphics[width=0.7\textwidth]{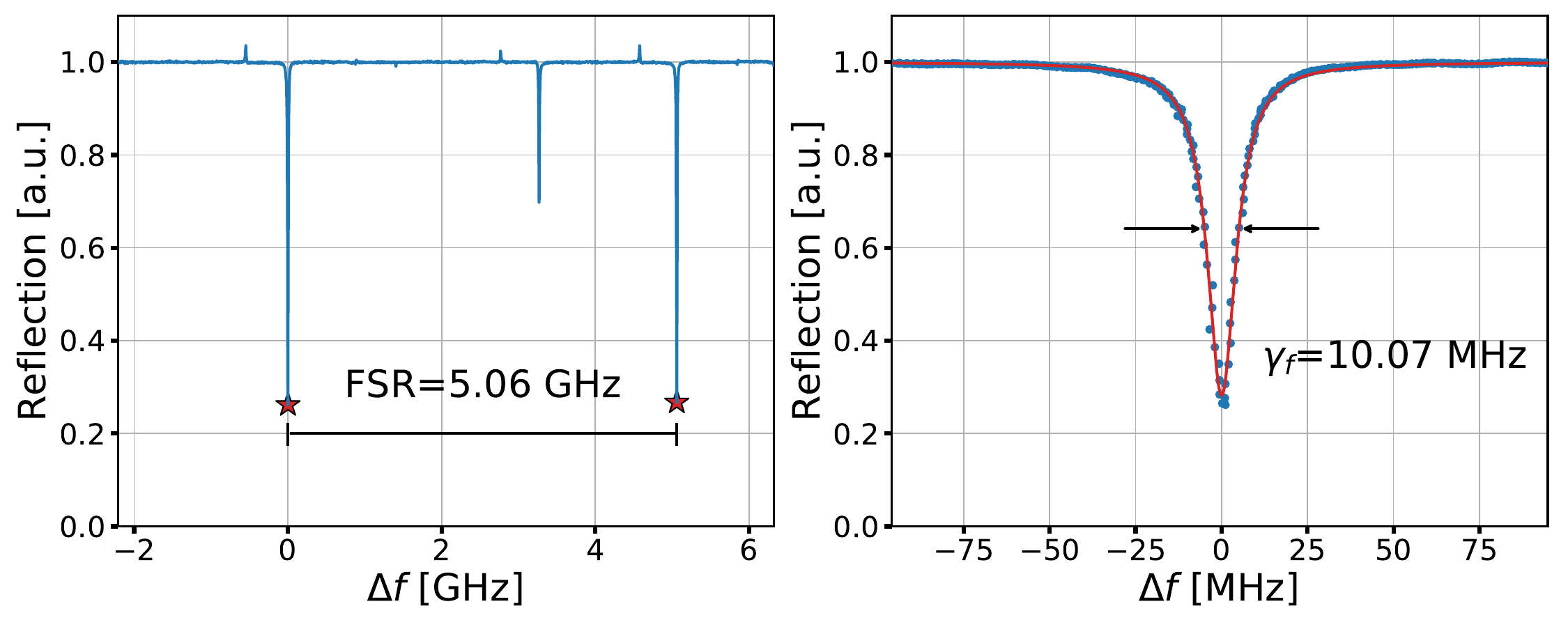}
    \caption{Reflection signal of one Fabry-Perot cavity as a function of the frequency difference $\Delta f$, which is proportional to the applied voltage on the piezoelectric actuator. The right graph is a zoom in of the left graph around the fundamental mode centered at $\Delta f=0$ (arbitrary). Blue points are experimental data and the red line is a Lorentzian fit. We measured $FRS=5.06$ GHz and $\gamma_f=10.07$ MHz.}
    \label{fig:Filters_carac}
\end{figure}

\newpage

\section{Pump-probe measurement and role of the average power}

We performed a pump-probe measurement to investigate the effect of the time delay between pulses. The optical signal sent to the disk consists of a repetition of a 50 $\mu$s sequence composed of a \textit{pump} pulse with spectral detuning $\delta_L=\Omega_m/2\pi$, duration $t_{pump}=4$ $\mu s$ and instantaneous power $P_{pump}=850$ $nW$, followed by a \textit{probe} pulse with spectral detuning $\delta_L=-\Omega_m/2\pi$, duration $t_{probe}=1$ $\mu s$ and instantaneous power $P_{probe}=85$ $nW$. The temporal delay $\Delta t$ between consecutive pump and probe pulses is varied while keeping the average power constant (hence, the temporal delay between consecutive probe and pump pulses is 45~$\mu$s $-\Delta t$). We show in Figure \ref{fig:mK_PP} the anti-Stokes optomechanically-scattered photons count rate at the optical resonance generated by the probe pulse as a function of the time delay $\Delta t$. This count rate is constant over all the covered time delays $\Delta t$ in the range 0.1 to 20 $\mu s$. Such behavior is expected if the thermalization time is much quicker than 0.1 $\mu s$ (compatible with disk thermal relaxation) or, on the contrary, if it is much larger than 20 $\mu s$ (compatible with slower extracavity heating processes).

\begin{figure}[!th]
    \centering
    \includegraphics[width=0.6\textwidth]{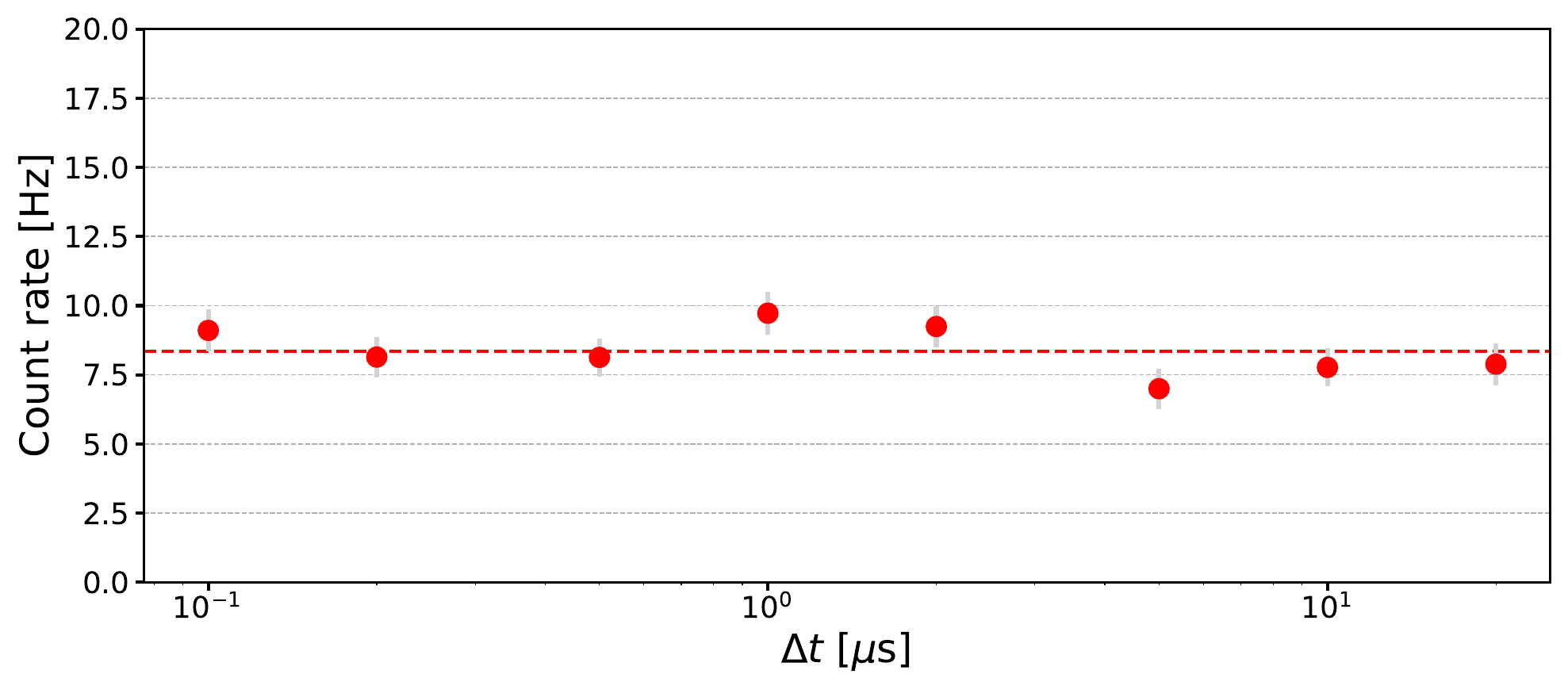}
    \caption{Anti-Stokes optomechanically-scattered photons count rate generated by the probe pulse as a function of the time delay $\Delta t$, once the dark count rate has been subtracted.}
    \label{fig:mK_PP}
\end{figure}

We also investigated the heating of the device induced by the average power of the optical sequence, while keeping the instantaneous power $P$ constant (and thus the intracavity photon number $\langle n_a\rangle$ constant). The average power is defined as $P_{avg}=P\cdot DC$, where $DC=\frac{\tau_p}{\tau+\tau_p}$ is the duty cycle of the optical pulse sequence with $\tau_p$ and $\tau$ are the pulse duration and delay between pulses, respectively. Varying $\tau$ allows varying $DC$ and thus $P_{avg}$. Figure \ref{fig:mK_sweepmeanpower} shows the red and blue count rates as a function of $P_{avg}$ when $\tau$ varies from 1 $\mu s$ to 96 $\mu s$. The variation of the count rate with $P_{avg}$ at constant $\langle n_a\rangle$ is an evidence of an extracavity heating (the substrate, its holder, the fibers) with a long thermalization time scale. 

\begin{figure}[!h]
    \centering
    \includegraphics[width=0.4\textwidth]{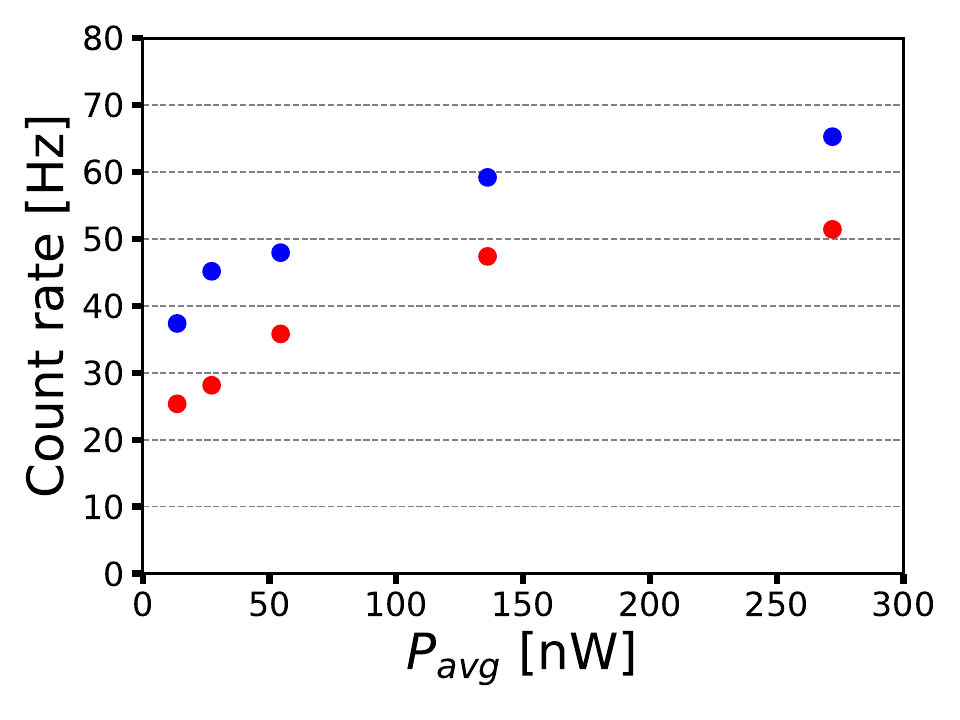}
    \caption{Anti-Stokes and Stokes optomechanically-scattered photons count rates generated by red and blue pulses once the dark count rate has been subtracted as a function of the optical mean input power $P_{avg}$, when $P\approx340$ nW.}
    \label{fig:mK_sweepmeanpower}
\end{figure}

\end{widetext}

\end{document}